\documentclass[preprint,12pt]{elsarticle}

\usepackage{graphicx}
\DeclareGraphicsExtensions{.pdf,.png}
\usepackage[colorlinks=true,bookmarks=false,citecolor=blue,
urlcolor=blue]{hyperref}

\usepackage{fancyhdr} 
\fancypagestyle{firststyle} 
{ \fancyhf{}
  \rfoot{}
  \lfoot{\tiny{\sffamily{NOTICE: this is the author's version of
        a work that was accepted for publication in the Journal of
        Quantitative Spectroscopy and Radiative Transfer. Changes
        resulting from the publishing process, such as peer review,
        editing, corrections, structural formatting, and other quality
        control mechanisms may not be reflected in this
        document. Changes may have been made to this work since it was
        submitted for publication. A definitive version was
        subsequently published in
        \href{http://dx.doi.org/10.1016/j.jqsrt.2012.06.007}{\textsl{Journal
            of Quantitative Spectroscopy and Radiative Transfer}
          \textbf{113}(18): 2482 (2012)}}}.}  
} 
\usepackage{amssymb}

\usepackage{url}

\journal{Journal of Quantitative Spectroscopy and Radiative Transfer}

\begin{document}

\begin{frontmatter}



  \title{Imaging Multiple Colloidal Particles by Fitting
      Electromagnetic Scattering Solutions to Digital Holograms}


\author[label1]{Jerome Fung}
\ead{fung@physics.harvard.edu}
\author[label2]{Rebecca W.~Perry}
\author[label1]{Thomas G.~Dimiduk}
\author[label2,label1]{Vinothan N.~Manoharan\corref{cor1}}
\ead{vnm@seas.harvard.edu}

\address[label1]{Dept.~of Physics, Harvard University, 17 Oxford Street, Cambridge, MA 02138, USA}
\address[label2]{School of Engineering \& Applied Sciences, Harvard University, 29 Oxford Street, Cambridge, MA 02138, USA}
\cortext[cor1]{Corresponding author}

\begin{abstract}
  Digital holographic microscopy is a fast three-dimensional (3D)
  imaging tool with many applications in soft matter physics. Recent
  studies have shown that electromagnetic scattering
    solutions can be fit to digital holograms to obtain the 3D
  positions of isolated colloidal spheres with nanometer precision and
  millisecond temporal resolution. Here we describe the results of new
  techniques that extend the range of systems that can be studied with
  fitting.  We show that an exact multisphere superposition scattering
  solution can fit holograms of colloidal clusters containing up to
  six spheres.  We also introduce an approximate and computationally
  simpler solution, Mie superposition, that is valid for multiple
  spheres spaced several wavelengths or more from one another. We show
  that this method can be used to analyze holograms of several spheres
  on an emulsion droplet, and we give a quantitative criterion for
  assessing its validity.
\end{abstract}


\end{frontmatter}


\section{Introduction}
\thispagestyle{firststyle}

\begin{figure}[p]
\centering
\includegraphics{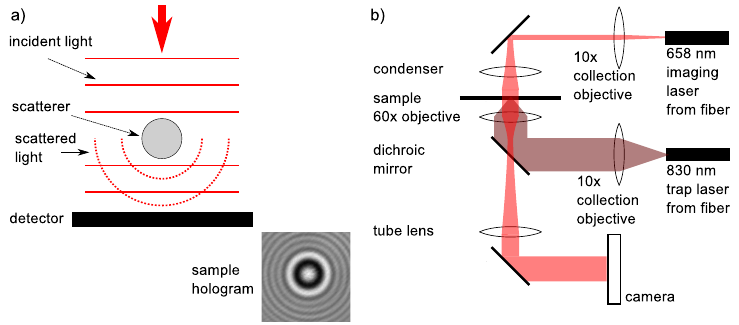}
\caption{\label{fig:Uberscope} a) Schematic illustration of in-line
  digital holographic microscopy. An incident plane wave illuminates a 
  scatterer, and a detector records the interference pattern formed by
  the scattered light and unscattered incident light. The inset shows
  a typical interference pattern, or hologram, formed by a single 
  1-$\mu$m-diameter polystyrene sphere in water. b) Schematic
  illustration of DHM optical train, showing the counterpropagating
  laser beams used for DHM (lighter red) and optical trapping (darker
  red).}
\end{figure}

\label{sec:intro}
In-line digital holographic microscopy (DHM) has emerged as a powerful
and useful 3-dimensional (3D) imaging technique for characterizing
colloids and soft materials \cite{lee_characterizing_2005,
  cheong_holographic_2008, cheong_technical_2009,
  xiao_multidimensional_2010, xiao_sorting_2010, fung_measuring_2011,
  kaz_physical_2012}.  Unlike conventional microscopes, holographic
microscopes use a coherent source to illuminate a sample, generating a
two-dimensional interference pattern, or hologram,
 that encodes 3D information
(Figure \ref{fig:Uberscope}a).  For relatively simple objects, such as
colloidal spheres, the most precise way to recover this 3D information
is to fit an electromagnetic scattering solution such as 
the Lorenz-Mie solution to the hologram. The fitting
technique can reveal the positions of particles with nanometer-scale
spatial precision \cite{lee_characterizing_2005} on sub-millisecond
time scales.  The high temporal resolution, more than an order of
magnitude larger than that of other 3D imaging techniques such as
confocal microscopy, has enabled observations of new physical
phenomena, as demonstrated in a recent study \cite{kaz_physical_2012}
showing the detailed dynamics of a colloidal particle interacting with
an oil-water interface.

Thus far, the fitting technique has been used 
only to analyze holograms of simple scatterers. The seminal
work of Lee \emph{et~al.} \cite{lee_characterizing_2005} as well as
recent experiments from Kaz \emph{et~al.} \cite{kaz_physical_2012}
examined single isolated spheres, for which the well-known Lorenz-Mie
solution suffices. Recently Fung \emph{et~al.} used a multisphere
superposition code \cite{mackowski_calculation_1996} to model 
digital holograms of non-rigid colloidal sphere dimers
and rigid trimers \cite{fung_measuring_2011}.  To date, however, there
have been no reports on using scattering solutions to analyze
holograms of more complex scatterers, including clusters of more than
3 particles or colloidal systems with configurations that are unknown
\textit{a priori}.

Such systems are of significant physical interest. Recently, Meng
\emph{et~al.} showed how clusters of six or more spheres with a
short-ranged attraction self-assemble into morphologies governed by
rotational and vibrational entropy
\cite{meng_free-energy_2010}. Detailed 3D observations of the
self-assembly process and transitions between different cluster
morphologies were not possible in that work, which relied on
conventional microscopy. Another interesting system consists of
colloidal particles bound to the surface of an emulsion droplet. The
particle configurations, which need not be rigid or symmetrical
\cite{mcgorty_measuring_2008}, may reveal information about the still
poorly-understood interaction forces between colloidal particles on a
curved liquid-liquid interface
\cite{nikolaides_electric-field-induced_2002}.

In this work we demonstrate new fitting methods for analyzing digital 
holograms of more complex scatterers, including rigid clusters
containing up to 6 particles and nonrigid configurations of particles
bound to the surface of an emulsion droplet. We fit two analytical
scattering solutions to our experimental data: an exact
multisphere superposition solution and an approximate, computationally
simpler approximation, Mie superposition. Both solutions can be used
to compute holograms from an arbitrary number of spherical particles
in an arbitrary geometry, but they differ in the precision to which
they can fit holograms of the two experimental systems.  We discuss
the applicability of the Mie superposition approximation and the
challenges remaining in the application of these techniques to other
colloidal systems.

\section{Experimental Methods}
\label{sec:methods}

\subsection{Sample Preparation}
\label{subsec:samples}

\begin{figure}[p]
\centering
\includegraphics[width=\textwidth]{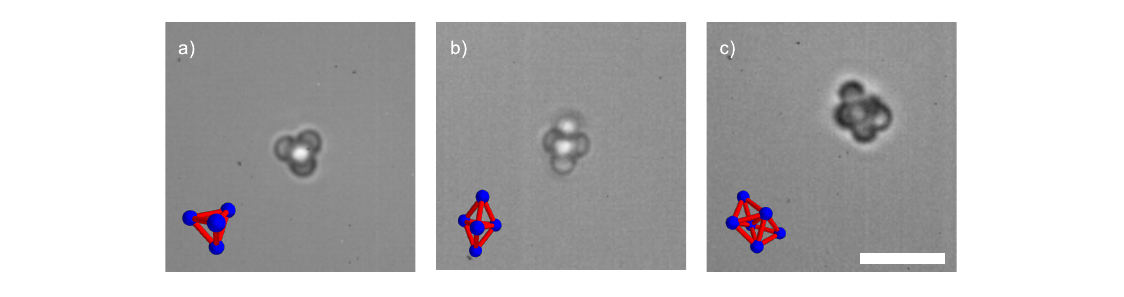}
\caption{\label{fig:brightfield} Bright field micrographs of clusters
  made from 1.3-$\mu$m diameter polystyrene spheres, recorded with a
  60$\times$ 1.20 NA water immersion objective. Ball-and-stick models,
  indicating cluster geometry and orientation, are a guide to the
  eye. Scale bar 5 $\mu$m. $n_s$ denotes the number of spheres in each
  cluster. a) Tetrahedron ($n_s=4$). b) Trigonal bipyramid
  ($n_s=5$). c) Polytetrahedron ($n_s=6$).}
\end{figure}

We image two different types of samples in this work: colloidal
clusters bound by depletion forces, and particles bound to the surface
of index-matched emulsion droplets.

\subsubsection{Colloidal Clusters}
We manually assemble tetrahedra containing four spheres, trigonal
bipyramids containing five spheres, and polytetrahedra containing six
spheres using the optical trap described in Section
\ref{sec:apparatus}. Bright-field micrographs of these clusters are
shown in Figure \ref{fig:brightfield}. To make the clusters, we first
prepare a dilute suspension of monodisperse, 1.3-$\mu$m-diameter
surfactant-free, sulfate-stabilized polystyrene spheres (Invitrogen)
at a volume fraction of $8\times 10^{-6}$ in an aqueous solution
containing 5 mM NaCl and 246 mM sodium dodecyl sulfate (SDS). We
load this suspension into sample cells made from untreated glass, as
described in \cite{fung_measuring_2011}. Because the SDS concentration
is far above the critical micelle concentration
\cite{mysels_critical_1971}, the SDS forms micelles which induce a
short-ranged depletion attraction between the polystyrene spheres
\cite{asakura_interaction_1958, vrij_polymers_1976}. This attraction
causes spheres to bind together into clusters when drawn into the
focus of the optical trap. In all cases, we verify the cluster
geometry by direct observation with bright-field microscopy before
holographic imaging.

\subsubsection{Particles on Emulsion Droplets}
\begin{figure}[p]
\centering
\includegraphics{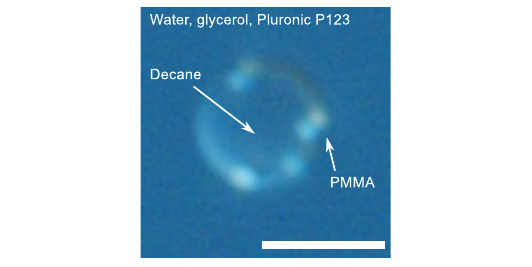}
\caption{\label{fig:ndep} Bright field micrograph of decane droplet
  with 4 PMMA spheres on its surface, recorded under 100$\times$
  magnification and differential interference contrast (DIC). Here, a
  slight index mismatch between the decane and the continuous phase
  makes the droplet visible under DIC. Scale bar 5 $\mu$m.}
\end{figure}
We prepare emulsion droplets laden with colloidal spheres by
dispersing a suspension of microspheres in oil into an aqueous
continuous phase. We suspend 0.8 $\mu$m diameter polymethyl
methacrylate (PMMA) spheres coated with poly(12-hydroxystearic acid)
stabilizer \cite{herzig_bicontinuous_2007, antl_preparation_1986} in
decane at a mass fraction of $2.1\times10^{-3}$. The aqueous
continuous phase contains 0.2\% w/w Pluronic P123 triblock copolymer
surfactant and 56\% w/w glycerol. The glycerol matches the refractive
index of the continuous phase to that of decane, so
that only the PMMA spheres scatter light. We prepare the emulsions
by mixing 0.5 mL of the PMMA-containing decane with 20 mL of the
continuous phase in a 40 mL scintillation vial and shearing the
mixture for 3 minutes at 9500 rpm with an Ika T9 Basic homogenizer
equipped with a S25N-8G dispersing tool. After emulsification, we
dilute the emulsion to 17\% v/v with additional continuous phase. A
micrograph of a typical emulsion droplet, laden with 4 particles but
with the continuous phase slightly mismatched to allow the droplet to
be seen, is shown in Figure \ref{fig:ndep}.

In order to minimize both the effects of nearby glass surfaces on the
interparticle interactions and unwanted back-reflections of the
scattered light during imaging, we use negative dielectrophoresis
(NDEP) to keep the particle-laden decane droplets away from the top of
our sample cells, to which they would otherwise rise due to
buoyancy. We prepare sample cells similar in design to those used for
the cluster measurements, except that the surface of the uppermost
glass microscope slide in contact with the emulsion contains a linear
array of interdigitated indium tin oxide (ITO) electrodes with a 40
$\mu$m spacing between adjacent electrodes
\cite{markx_dielectrophoretic_1997, huang_introducing_1997,
  morgan_dielectrophoretic_2001}. By applying a 10 V peak-to-peak, 300
kHz square wave with an arbitrary waveform generator (Agilent AFG
3022B), we produce a downward NDEP force on the droplets. Diffraction
from the edges of the ITO electrodes is negligible.

We use photolithography and wet etching to prepare the interdigitated
electrode arrays on microscope slides coated with a 30 nm layer of ITO
(Delta Technologies, CB-90IN coating).  We spin-coat the ITO surface
of the slides with Shipley S1813 positive photoresist at 5000 rpm and
soft bake the slides on a 115$^\circ$C hot plate for 1 minute. We
first define the electrode pattern by exposing the photoresist to UV
light through a photomask (150 mJ/cm$^2$ exposure at 405 nm), then
develop the photoresist by immersion in Microposit MF CD-26
developer for 1 minute at room temperature. Following an overnight
hard bake in a 90$^\circ$C oven, we etch away the exposed ITO with
an aqueous solution containing 40\% v/v HCl and 10\% v/v HNO$_3$ for
12 minutes at room temperature.

\subsection{Recording Digital Holograms}
\label{sec:apparatus}

We record digital holograms on a modified Nikon inverted microscope
(Figure \ref{fig:Uberscope}b). Our apparatus is described in detail in
\cite{fung_measuring_2011, kaz_physical_2012}. In brief, we couple a
660 nm laser diode (Opnext HL6545MG) into a single-mode fiber, from
which we obtain up to 60 mW of laser power. The beam is collimated
by the microscope's condenser before striking the sample. To image
colloidal clusters, we use a 60$\times$, 1.20 NA water immersion
objective (Nikon). To image particles on droplets, we use a
100$\times$, 1.40 NA oil immersion objective (Nikon) with an immersion
oil of index 1.412 (Cargille).  In both cases, the index of refraction
of the immersion medium is chosen to match that of the continuous
phase of the sample, thus minimizing height-dependent spherical
aberration \cite{biancaniello_line_2006}.  Images are recorded by a
Photon Focus MVD-1024E-160 camera and captured to disk by an EPIX
PIXCI E4 frame grabber.

We also integrate an optical trap into the apparatus to manually
assemble colloidal clusters and manipulate them. The trap beam is
kept off during holographic imaging. Light from the 830 nm trap diode
(Sanyo DL-8142-201) is filtered through a single-mode fiber and
reflected by a dichroic mirror into the back aperture of the Nikon
objective lens used for imaging, as shown in Figure
\ref{fig:Uberscope}b.

\section{Fitting Scattering Solutions to Digital Holograms}
\subsection{Pre-Processing Holograms \& Fitting Scattering Models}

After recording digital holograms and suitable backgrounds in an
experiment, we fit scattering models to the holograms.  We have
released our software tools for doing so in the open-source package
Holopy, available at
\href{https://launchpad.net/holopy}{https://launchpad.net/holopy}.  As
described in detail in \cite{fung_measuring_2011}, we divide the
recorded holograms by a background taken in an empty field of view and
normalize them to a mean value of 1. We then use the
Levenberg-Marquardt algorithm to minimize the sum of residuals between
the experimental holograms and model holograms computed from a set of
fixed and variable physical parameters, including particle
coordinates, sizes, and refractive indices. We discuss the models in
more detail in Section \ref{sec:model}.

Two features of the Levenberg-Marquardt iterative minimization algorithm 
have significant implications for our analysis. First, the algorithm 
requires initial values for all the model parameters. Typically, we 
choose these initial values to provide a qualitative match to the 
experimentally recorded holograms, allowing the algorithm to converge
after approximately 10 iterations. Second, the algorithm requires the
number of spheres in the hologram to be known \textit{a priori}. When
necessary, we can determine the number of spheres using complementary 
hologram analysis techniques, such as Rayleigh-Sommerfeld reconstruction
\cite{kreis_frequency_2002}.

We choose the parameters in the scattering models to obtain as much
physical information from the holograms as possible while minimizing
the required computational time.
When modeling holograms of rigid clusters, we assume that all of the
spheres in the clusters have the same radius and real refractive index,
and that the gap between spheres is uniform. We vary the common
radius, the common particle refractive index, the common gap distance,
the cluster center of mass, the three orientational Euler angles, and
the scaling coefficient $\alpha$ (see equation
\ref{eq:hologram}). When modeling holograms of particles on an emulsion
droplet, we vary the common particle radius, the common particle
refractive index, 3 spatial coordinates per particle, and the scaling
coefficient $\alpha$. 
While we could use models with more
parameters, such as arbitrary radii for all the spheres, we gain little
useful information from doing so, since our spheres are monodisperse.
Fitting more complex models requires significantly more computation time, 
since the time required to fit a model with $N_{params}$ parameters scales 
as $N_{params}^2$.

\subsection{Modeling Holograms}
\label{sec:model}
In our models, the normalized hologram $I_{hol}$ may be described as
\begin{equation} \label{eq:hologram}
I_{hol} = 1 + 2\alpha\Re \left[ \mathbf{E}_{scat} \cdot \mathbf{\hat{e}} \right]
 + \alpha^2|\mathbf{E}_{scat}|^2
\end{equation}
where $\mathbf{E}_{scat}$ is the scattered electric field, and
$\mathbf{\hat{e}}$ is the unit vector describing the incident
polarization. As described in \cite{lee_characterizing_2005,
  fung_measuring_2011}, $\alpha$ is inversely proportional to the
magnitude of the incident electric field $\mathbf{E}_{inc}$ and allows
us to model variations in the incident laser intensity. In our
experiments, we know the optical parameters of the system, including
the medium refractive index, the incident laser wavelength, and the
apparent detector pixel size (after magnification). Therefore, 
we require only an algorithm for computing $\mathbf{E}_{scat}$
to model a hologram of some given configuration of particles.
Any such algorithm must include the full radial dependence of
$\mathbf{E}_{scat}$ on spherical Hankel functions of the first kind
and their derivatives, as the far-field asymptotic approximation does
not hold for the distances at which we record holograms
\cite{fung_measuring_2011}.

We use two analytical methods to calculate $\mathbf{E}_{scat}$ for
multiple spheres. The first method, described in detail in
\cite{fung_measuring_2011}, uses the multisphere superposition code
SCSMFO developed by Mackowski \emph{et~al.}
\cite{mackowski_calculation_1996}. Previously we termed this approach
the ``$T$-matrix method'', but we now prefer the term ``multisphere
superposition'' since we never actually compute the $T$-matrix itself,
which is typically used for orientational averaging. 
This method
gives exact numerical solutions to Maxwell's equations for
the scattering of a plane wave by multiple 
spheres, accounting for all electromagnetic coupling
between the particles \cite{mackowski_calculation_1996,
  mishchenko_t-matrix_1996}.

The second approach is to superpose the
fields calculated from the Lorenz-Mie solution for each of the
spheres, taking into account the phase differences arising from the
displacement of the spheres along the optical axis. This approach,
which we term Mie superposition, assumes that only a plane wave falls
onto each sphere. 
It neglects electromagnetic coupling, 
including multiple scattering, between the spheres.
$\mathbf{E}_{scat}$ calculated from Mie superposition is the
lowest-order approximation to multisphere superposition; it is
the initial solution that SCSMFO iteratively corrects in the
order-of-scattering approach \cite{mishchenko_light_2000}.

The main advantage of Mie superposition over multisphere superposition
is that it is less computationally costly for multiple spheres that
are spaced far apart. SCSMFO
expresses $\mathbf{E}_{scat}$ as a superposition of vector spherical
harmonics whose origin lay at the center of mass of all the
particles. It does so using vector spherical harmonic translation
theorems to translate field expansions centered about each sphere to
the cluster center of mass \cite{mackowski_calculation_1996}. When the
particles are far from their center of mass, the final field expansion
contains many more terms than would be necessary in a field
expansion centered about the individual spheres. We therefore use the
Mie superposition approach for our submicron-diameter PMMA particles
on emulsion droplets, which are weakly scattering (size parameter
$x\approx5$ and relative index $m\approx1.07$) and spaced far
apart. We discuss the validity of Mie superposition further in Section
\ref{sec:ms_v_tm}.

After fitting a scattering model to a hologram, we quantitatively
confirm the fit of the model by examining two statistical
measures of the goodness-of-fit. The first measure, 
chi-squared per pixel $\chi^2_p$, is the quantity the fitting algorithm
attempts to minimize:
\begin{equation}
\chi^2_p = \frac{1}{N}\sum_{i=1}^N (I_{holo} - I_{fit})^2.
\end{equation}
The sums run over all $N$ pixels of the recorded normalized hologram
$I_{holo}$ and the best-fit model hologram $I_{fit}$. For any given
hologram, comparing $\chi^2_p$ to an expected noise level allows us to
assess whether deviations between the recorded hologram and best-fit
model are due to instrumental noise or to a systematic error in the
model's description of the underlying data. Assuming noise in the
least significant bit of an 8-bit camera, we would expect $\chi_p^2$
values greater than $(1/255)^2 = 1.54\times10^{-5}$ to originate
from systematic errors. 

The second statistical measure we use is the coefficient of
determination $R^2$.  We define $R^2$ as
\begin{equation}
  R^2 = 1 - \frac{\sum_{i=1}^N (I_{holo} - I_{fit})^2}{\sum_{i=1}^N (I_{holo} - \bar{I}_{holo})^2} = 1 - \frac{\sum_{i=1}^N (I_{holo} - I_{fit})^2}{\sum_{i=1}^N (I_{holo} - 1)^2}
\end{equation}
where $\bar{I}_{holo}$ is the mean value of the recorded hologram,
which is 1 by our normalization
\cite{mendenhall_statistics_1991}. $R^2$ measures the fraction of the
variation of the recorded hologram from its mean value that is
captured by the best-fit model, independent of the amount of variation
in the hologram. 
Whereas $\chi^2_p$ varies significantly across
physical systems that differ in scattering cross section and hence 
hologram fringe amplitude, $R^2$ does not.  Therefore, we use $R^2$ 
to assess the validity of the scattering models fitted to the 
holograms. In particular, as we discuss in Section \ref{sec:ms_v_tm},
$R^2$ helps to assess the validity of the Mie superposition
approximation.

\section{Results and Discussion}
\label{sec:results}

\subsection{Rigid Clusters}
\label{sec:clusters}
\begin{figure}[p]
\centering
\includegraphics[width=\textwidth]{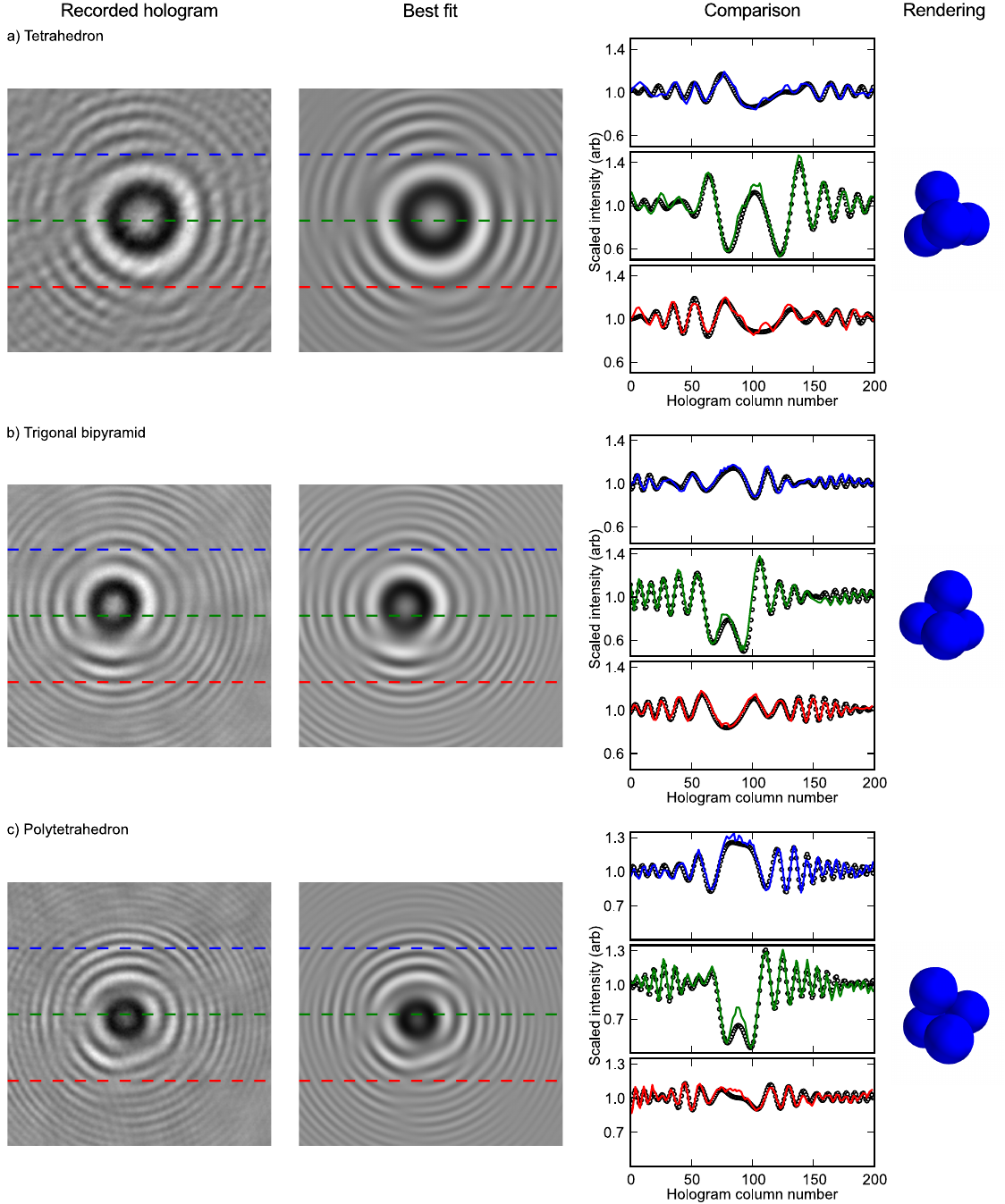} 
\caption{\label{fig:cluster_montage} Recorded and best-fit model
  holograms of rigid clusters. The third column compares the recorded
  holograms (solid lines) to the best-fit models (open symbols) along
  the color-coded dashed horizontal lines shown in the holograms. The
  renderings in the rightmost column show the cluster orientations
  determined from the fits. In the renderings, the incident light
  propagates into the page.  a) Tetrahedron ($n_s = 4$). b) Trigonal
  bipyramid ($n_s = 5$). c) Polytetrahedron ($n_s = 6$).}
\end{figure}

\begin{table}[p]
\centering
\begin{tabular} {l c c c}
Cluster & Radius (nm) &  $\chi^2_p$ & $R^2$\\
\hline
Tetrahedron & $670\pm30$ &  $1.18\times 10^{-3}$ & 0.923\\
Trigonal bipyramid &  $640\pm30$ & $9.48\times 10^{-4}$ & 0.910\\
Polytetrahedron & $650\pm20$  & $1.24 \times 10^{-3}$ & 0.877 \\
\end{tabular}
\caption{\label{tab:rigid_data} Fitted radii and goodness-of-fit
statistics  $\chi^2_p$ and $R^2$ for rigid cluster holograms in Figure 
\ref{fig:cluster_montage}.}
\end{table}

Figure \ref{fig:cluster_montage} compares recorded and best-fit model
holograms, calculated with multisphere superposition, for tetrahedral,
trigonal bipyramidal, and polytetrahedral clusters.  Qualitatively, we
observe excellent agreement between the recorded and best-fit
holograms. In particular, the best-fit models reproduce the highly
non-axisymmetric fringes in the recorded holograms, which depend strongly
on the cluster orientations. The quality of the agreement is
confirmed by the $R^2$ values of the fits, which are close to 1 (Table 
\ref{tab:rigid_data}). Also, the fitted particle radii are close to 
the manufacturer's reported value of 650 nm.  However,
the values of $\chi^2_p$ we observe are an order of magnitude larger
than what we would expect due to camera noise, and indicate that further
improvements to fits will depend on modeling additional physical phenomena or
improving the convergence of the fitter.

\subsection{Particles on Droplets}
\label{sec:particles_on_drops}
\begin{figure}[p]
\centering
\includegraphics[width=\textwidth]{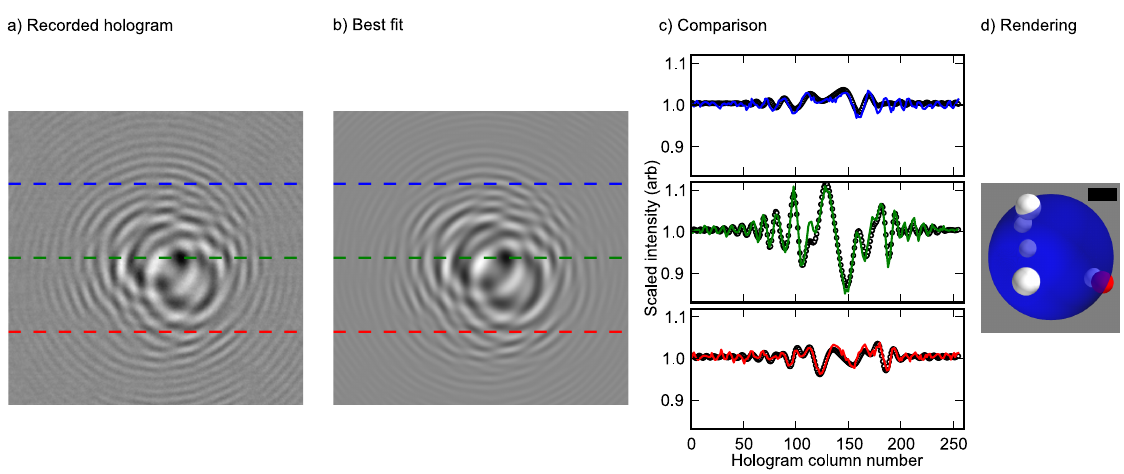}
\caption{\label{fig:dropholos}Holograms of six PMMA spheres on a
  decane droplet. a) Recorded hologram. b) Best-fit model. c)
  Comparison between recorded hologram (solid lines) and best-fit
  model computed using Mie superposition (open symbols) along the 
  dashed horizontal lines in the
  holograms.  d) Rendering showing the sphere positions determined by
  fitting the holograms. The incident light propagates into the page,
  and the scale bar is 1 $\mu$m. The larger blue sphere is a guide to
  the eye; its position and diameter, 4.35 $\mu$m, were determined by
  fitting a sphere to the coordinates of the six particles. The small
  red sphere indicates the particle showing the largest discrepancy in
  position along the optical axis between fits to Mie superposition 
  and multisphere superposition.}
\end{figure}
Here we show that our fitting techniques may also be applied to
multisphere systems without a fixed geometry. Figure
\ref{fig:dropholos} shows a hologram of six particles bound to the
surface of a decane droplet and a best-fit model calculated using Mie
superposition. Again, the qualitative agreement between the fringes of
the recorded hologram and the best-fit model is good.  Quantitatively,
we find $\chi^2_p = 7.29\times10^{-5}$ and $R^2 = 0.811$. $\chi^2_p$
is much lower than the values we obtained for the clusters, primarily
because the peak amplitude of the hologram in Figure
\ref{fig:dropholos} is significantly smaller than the peak amplitudes
of the cluster holograms in Figure \ref{fig:cluster_montage}. The
value of $R^2$ indicates that the fit is slightly worse than the fits
for the clusters.  However, because we know the particles are bound to
the surface of a spherical droplet, we can independently test the
accuracy of the fitted particle positions. While we cannot directly
image the decane droplet, which is index-matched to the continuous
phase, 
we can fit the surface of a sphere to the particle coordinates,
as shown in Figure \ref{fig:dropholos}d. The average difference
between the radial distance of each particle from the droplet center
and the fitted droplet radius is $60\pm60$ nm. Differences of this
scale are comparable to previously reported precisions for DHM
\cite{lee_characterizing_2005, fung_measuring_2011}, and may be
partially accounted for by variations in the interfacial contact angle
between different particles \cite{kaz_physical_2012}.

To determine whether the slightly worse value of $R^2$
obtained in this fit is due to the Mie superposition approximation, we
examine the validity of this approximation in further detail.

\subsection{Applicability of Mie Superposition}
\label{sec:ms_v_tm}
To confirm that Mie superposition is a suitable means for analyzing
holograms like that in Figure \ref{fig:dropholos}, in which multiple
weakly scattering spheres are situated several diameters apart, we fit
a model based on multisphere superposition to the same hologram. The
multisphere superposition fit yields $\chi^2_p = 8.08\times10^{-5}$
and $R^2 = 0.790$, comparable to the Mie superposition
fits. Differences between fitted particle coordinates in the two
in-plane directions, perpendicular to the optical axis, are at most 22
nm.  The largest difference between fitted coordinates along the
optical axis is 154 nm. The sphere showing the largest displacement
along the optical axis is highlighted in red in Figure
\ref{fig:dropholos}d.  The size of these differences, as well as the
lack of improvement in the goodness-of-fit using the multisphere
superposition code, indicate that Mie superposition is an appropriate
approximation. We conclude that the smaller $R^2$ values for this
system stem from physical effects---such as weak scattering by the
decane droplet---that neither multisphere superposition nor Mie
superposition account for.

Further insight into the applicability of Mie superposition comes from
examining the particle showing the largest coordinate difference along
the optical axis. As shown in Figure \ref{fig:dropholos}d, the largest
difference occurs when two particles nearly occlude one another. In
such a configuration, the assumption that the field incident on
each sphere is simply the illuminating plane wave is clearly
invalid, as the colloidal spheres scatter most strongly in the forward
direction. The field incident on the occluded sphere should therefore
include a significant component of the scattered field from the first
sphere. Whereas the multisphere superposition solution accounts for
this multiple scattering effect, Mie superposition does not.

\begin{figure}[pt]
\centering
\includegraphics[width=\textwidth]{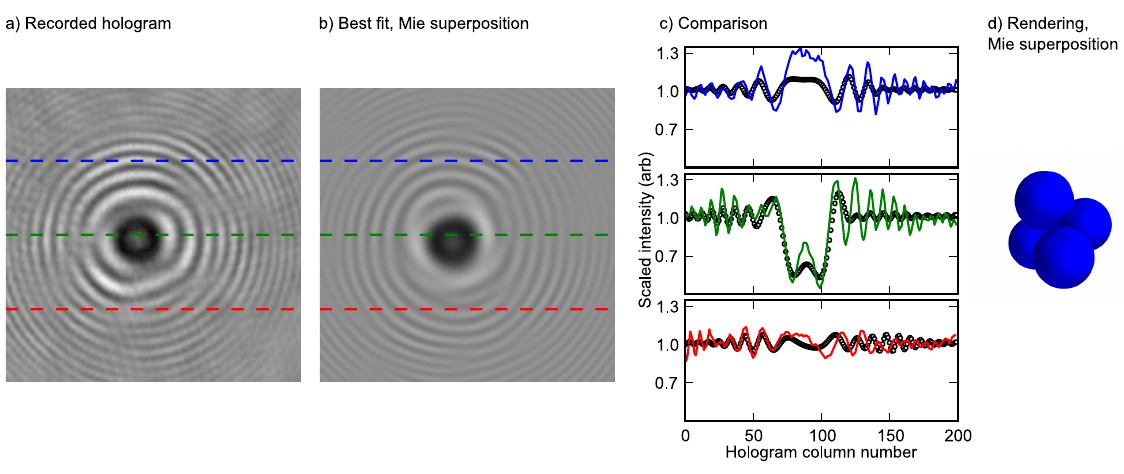}
\caption{\label{fig:ms_polytet} Fit of polytetrahedron hologram in
  Figure \ref{fig:cluster_montage}c performed using Mie
  superposition. a) Recorded hologram (shown again for ease of
  comparison). b) Best-fit model determined from Mie superposition. c)
  Comparison between recorded hologram (solid lines) and Mie
  superposition model (open symbols) along the dashed lines in the
  holograms. d) Rendering showing cluster orientation determined by
  Mie superposition fit. The incident light propagates into the page.}
\end{figure}

As expected, Mie superposition fails to fit holograms from the
clusters discussed in Section \ref{sec:clusters}, where the
constituent spheres have a relative index $m\approx 1.2$ and size
parameter $x\approx 8.3$.  A fit to the polytetrahedron hologram of
Figure \ref{fig:cluster_montage}c using a Mie superposition model
yields $\chi_p^2 = 5.39\times10^{-3}$ and $R^2=0.463$ (Figure
\ref{fig:ms_polytet}). These values are much poorer than the values 
associated with the multisphere superposition model. Moreover, 
qualitative differences between the best-fit Mie superposition model
and the experimental hologram are readily apparent.

To quantify the validity of Mie superposition for calculating
holograms---or any other scattering-related physical quantity---of
multiple spheres, we propose a dimensionless figure of merit. Mie
superposition assumes that the exciting field at any sphere,
$\mathbf{E}_{ex}$, is approximately equal to the incident plane wave
$\mathbf{E}_{inc}$. Following the multisphere superposition approach,
consider $\mathbf{E}_{ex}$ for any given sphere $i$ to be equal to the
sum of the incident plane wave and the scattered waves from every
other sphere at $i$:
\begin{equation}
\mathbf{E}_{ex,i} = \mathbf{E}_{inc} + \sum_{j\ne i}^N \mathbf{E}_{scat,j}.
\end{equation}
For Mie superposition to be valid, $|\mathbf{E}_{scat,j}|$ must be
much smaller than $|\mathbf{E}_{inc}|$. This requires the particles to
be far enough apart that their near fields do not couple. Then,
$|\mathbf{E}_{scat,j}|$ scales approximately as
$|\mathbf{E}_{inc}|S/kR$, where $R$ is a typical interparticle
distance and $S$ denotes the magnitude of the amplitude scattering
matrix of sphere $j$ in the Mie solution.  From the optical theorem,
in the Mie solution the extinction efficiency $Q_{ext}$ for a single
sphere is given by
\begin{equation}
Q_{ext} = \frac{4}{x^2} \Re\left[ S(0) \right]
\end{equation}
where $S(0)$ is the complex amplitude scattering matrix evaluated in
the forward direction \cite{bohren_absorption_1998}.  Then, if
$Q_{ext}x^2/kR \ll 1$, $|\mathbf{E}_{scat,j}| \ll |\mathbf{E}_{inc}|$, 
and Mie superposition should be valid.

Taking the fitted drop diameter of 4.35 $\mu$m as a typical
interparticle spacing for PMMA spheres like those in the droplet
experiments, we find $Q_{ext}x^2/kR = 0.11$, in agreement with our
previous conclusion that the Mie superposition approximation is
accurate for this sample.  However, if we consider the sphere shown in
red in Figure \ref{fig:dropholos}d, using the
nearest-neighbor distance of 1.19 $\mu$m for $R$ yields $Q_{ext}x^2/kR
= 0.35$. This larger value indicates that Mie superposition is a
poorer approximation for this particle, as borne out by the 150 nm
difference in the fitted position from Mie superposition and
multisphere superposition. In contrast, for the polystyrene spheres
used in the cluster experiments, which are separated from each other
by approximately a particle diameter, $Q_{ext}x^2/kR = 13$. We
therefore conclude that $Q_{ext}x^2/kR$ should be approximately $0.1$
or smaller for Mie superposition results to be trusted to a precision
of $100$ nm or better.

When $Q_{ext}x^2/kR\ll1$ and $kR$ is large, Mie superposition is not
only valid but advantageous compared to multisphere superposition.  On
a desktop equipped with a 3 GHz Intel Core Duo processor, computing a
$256\times256$ hologram for the configuration of particles illustrated
in Figure \ref{fig:dropholos}d requires 11.8 seconds for multisphere
superposition and 3.0 seconds for Mie superposition. In contrast,
computing the $200\times200$ hologram of the six-particle
polytetrahedron of Figure \ref{fig:cluster_montage}c requires 3.9
seconds for multisphere superposition and 1.8 seconds for Mie
superposition. Here multisphere superposition is only 2 times slower,
an acceptable penalty given that the Mie superposition fit fails.

\subsection{Improving the Fits}
Our values of $\chi^2_p$ are larger than the expected noise
level. This indicates that our modeling and fitting procedures, rather
than instrumental noise, limit the goodness of our fits. Challenges
for future work fall into two categories: modeling additional physical
phenomena, and properly optimizing model parameters to determine a
best fit.

Our current hologram models do not capture several physical effects.
First, while we choose immersion fluids for our microscope objective
lenses to index-match the continuous phases of our samples as closely
as possible, we cannot completely eliminate spherical aberration.
Second, although we have fit for only a single, average particle
radius in these experiments, no colloidal suspension is perfectly
monodisperse.  
Third, while the well depth of the depletion attraction in the
clusters is at least several $k_BT$, thermally-induced vibrations
can perturb the cluster structures.  We are currently working to
model and account for spherical aberration in our hologram
analysis. Fitting for individual particle positions in a cluster,
rather than assuming geometric regularity, may address the latter two
phenomena.

Local minima in the fitting landscape pose another challenge to
obtaining good fits. The number of local minima may increase with the
number of parameters in the hologram models.  This is partly why we
constrain all the spheres in our scattering models to be the same
size. However, incorporating additional physical parameters into our
hologram models will be necessary to study more complex scatterers,
such as nonrigid clusters of weakly interacting spheres. Consequently,
in order to prevent fits from becoming trapped in local minima, we are
currently exploring the use of other minimizers, including one based
on the $r$-algorithm \cite{shor_minimization_1974}.

\section{Conclusions}
We have demonstrated that superposition solutions are accurate models
for fitting digital holograms of up to six colloidal spheres.  The
multisphere superposition solution can reproduce the holograms of
clusters in which there is significant coupling of the near-fields
between spheres.  We have also introduced a faster computational tool,
Mie superposition, that should facilitate the study of multisphere
systems where the particles are far apart.  This approximate technique
is accurate when $Q_{ext}x^2/kR\ll1$.

The six-particle clusters we have imaged are large enough to show
non-trivial dynamics related to multiple ground states
\cite{meng_free-energy_2010}.  We have also imaged configurations of
spheres without a fixed geometry.  Our experiments are thus the first
step towards measuring internal rearrangements and self-assembly of
sphere clusters. Moreover, analysis of dynamical information obtained
from time series of holograms like those studied in Figure
\ref{fig:dropholos} may yield insight into colloidal interactions at
liquid-liquid interfaces.

\section*{Acknowledgments}
We thank Andrew B.~Schofield for synthesizing PMMA microspheres, and
Keith A. Brown for assistance with the fabrication of NDEP
devices. Rebecca W.~Perry and Thomas G.~Dimiduk acknowledge the
support of National Science Foundation (NSF) Graduate Research 
Fellowships. This work was
funded by the NSF through CAREER grant no.~CBET-0747625 and through
the Harvard MRSEC, grant no.~DMR-0820484. Computations were performed
on the Odyssey cluster, managed by the Harvard FAS Sciences Division
Research Computing Group. NDEP device fabrication was performed at the
Center for Nanoscale Systems (CNS), a member of the National
Nanotechnology Infrastructure Network (NNIN), which is supported by
the NSF under grant no.~ECS-0335765. CNS is part of Harvard
University.






\bibliographystyle{elsarticle-num}
\bibliography{fung}







\end{document}